\documentclass{article}
\usepackage{bm}

\usepackage[preprint, nonatbib]{neurips_2024}

\usepackage[utf8]{inputenc} 
\usepackage[T1]{fontenc}    
\usepackage{hyperref}       
\usepackage{url}            
\usepackage{booktabs}       
\usepackage{amsfonts}       
\usepackage{nicefrac}       
\usepackage{microtype}      
\usepackage{xcolor}         
\usepackage{amsmath}
\usepackage{graphicx}
\usepackage{amssymb}
\usepackage{tabularx}
\usepackage{booktabs}
\usepackage{subcaption}
\usepackage{algorithm}
\usepackage{algorithmicx}
\usepackage{algpseudocode}

\title{Lattice Protein Folding with Variational Annealing}

\author{%
  Shoummo A. Khandoker\\
  Department of Computer Science, Indiana University Bloomington, \textit{Bloomington, IN 47405, USA} \\
  \texttt{sakhando@iu.edu} \\
  \And
  Estelle M. Inack \\
    Perimeter Institute for Theoretical Physics, \textit{Waterloo, Ontario, Canada}\\
    yiyaniQ, \textit{Toronto, Ontario, Canada}\\
    Department of Physics and Astronomy, University of Waterloo, \textit{Waterloo, Ontario, Canada}\\
  \texttt{einack@perimeterinstitute.ca} \\
  \And
  Mohamed Hibat-Allah \\
    Department of Applied Mathematics, University of Waterloo, \textit{Waterloo, Ontario, Canada}\\
    Vector Institute, \textit{Toronto, Ontario, Canada}\\
    Perimeter Institute for Theoretical Physics, \textit{Waterloo, Ontario, Canada}\\
  \texttt{mhibatallah@uwaterloo.ca} \\
}

\begin{document}

\maketitle

\begin{abstract}
Understanding the principles of protein folding is a cornerstone of computational biology, with implications for drug design, bioengineering, and the understanding of fundamental biological processes. Lattice protein folding models offer a simplified yet powerful framework for studying the complexities of protein folding, enabling the exploration of energetically optimal folds under constrained conditions. However, finding these optimal folds is a computationally challenging combinatorial optimization problem. In this work, we introduce a novel upper-bound training scheme that employs masking to identify the lowest-energy folds in two-dimensional Hydrophobic-Polar (HP) lattice protein folding. By leveraging Dilated Recurrent Neural Networks (RNNs) integrated with an annealing process driven by temperature-like fluctuations, our method accurately predicts optimal folds for benchmark systems of up to 60 beads. Our approach also effectively masks invalid folds from being sampled without compromising the autoregressive sampling properties of RNNs. This scheme is generalizable to three spatial dimensions and can be extended to lattice protein models with larger alphabets. Our findings emphasize the potential of advanced machine learning techniques in tackling complex protein folding problems and a broader class of constrained combinatorial optimization challenges.
\end{abstract}

\section{Introduction \& Previous Work}
Protein folding is a biological process in which a linear sequence of amino acids adopts a three-dimensional structure. A correct fold or a misfold can significantly affect the biological health of a living organism~\cite{folding_perspective}. As a result, an accurate understanding of how proteins fold is critical in biology and drug discovery~\cite{dill_protein_2008}. The curse of dimensionality of the protein folding space makes it challenging to address using standard computer simulations~\cite{somuchmoretoknow}. Lattice protein folding provides a simplified yet insightful framework for studying protein folding dynamics by reducing the complexity of the search space. In the regular lattice, each cell may house an amino acid. It is also common to further simplify this folding process by reducing all 20 types of amino acids to only two types: hydrophobic and polar amino acids, also called beads. These simplifications correspond to the Hydrophobic-polar (HP) lattice protein folding model~\cite{dill_theory_1985}, which is a model that helps elucidate fundamental principles of protein folding, such as the role of hydrophobic interactions and the relationship between amino acid sequences and native structures~\cite{lau1989lattice}. Despite these simplifications, finding the fold with the lowest energy (global minima) is NP-complete for both 2D and 3D HP lattice models \cite{lau1989lattice}.

Machine learning tools have already addressed the question of protein folding in different settings. In the continuous folding space, AlphaFold, a machine learning-based approach has achieved remarkable performance on the prediction of protein structures compared to state-of-the-art methods~\cite{jumper_highly_2021}. In the discrete folding space, machine learning approaches have also addressed the HP protein folding model in 2D as a combinatorial optimization problem. In recent years, there has been increasing interest in using machine learning techniques to solve combinatorial optimization problems~\cite{bengio2020machinelearningcombinatorialoptimization}. These techniques span different paradigms of learning such as reinforcement learning~\cite{bello2017neural,bresson2021transformernetworktravelingsalesman} and unsupervised learning~\cite{VNA2021, karalias2020erdos}. They also incorporate a wide variety of neural network architectures such as recurrent neural networks~\cite{Khandoker_2023, VNA2021}, graph neural networks~\cite{sanokowski2023variationalannealinggraphscombinatorial, sun2022annealed}, and diffusion models~\cite{sanokowski2024diffusionmodelframeworkunsupervised}. In particular, for the 2D lattice protein folding problem, this study~\cite{li2018foldingzeroproteinfoldingscratch} developed a model called FoldingZero, which combines deep reinforcement learning (RL) with a two-head deep convolutional neural network (HPNet) and a modified tree search algorithm. This work investigated chain sizes up to $85$ with successful runs matching the optimal energy for sizes up to $20$. Another RL study~\cite{Yu2020DeepRL} also investigated HP chains up to $36$ beads using various RL methods such as policy and value iteration, Monte Carlo Tree Search, and AlphaGo Zero with pretraining. According to their results, the adoption of the AlphaGo Zero algorithm exhibits superior performance in comparison to the other methods. Finally, by far the strongest RL work~\cite{Yang_2023} obtains optimal folds for HP chains up to $50$ beads. They largely attribute the success of their method to the incorporation of Long Short Term Memory (LSTM) architectures in their procedure, which capture long-range interactions in the folding process.

Our work demonstrates the ability of dilated recurrent neural networks (RNNs)~\cite{chang2017dilated} supplemented with temperature annealing ~\cite{VNA2021, sun2022annealed} to solve instances of the 2D HP lattice folding model. In all the previous studies, sampling invalid folds has been discouraged by introducing an energy penalty. In our work, we use masking to sample valid folds from RNNs autoregressively to enhance convergence and training stability. We also introduce a novel scheme by introducing a free energy upper bound that stabilizes and enhances RNNs training on the 2D HP model, while preserving their ability to generate folds in the valid folding space autoregressively. 

The plan of this paper is as follows: In the methods section, we describe the mathematical details of the HP model and the variational annealing framework that we use in conjunction with dilated RNNs. Additionally, we present our scheme for projecting dilated RNNs autoregressive sampling to valid folds and show our derived upper bound training which enhances the trainability of RNNs. Finally, in the Results and Discussion section, we highlight empirical evidence in favor of annealing and upper-bound training. We also highlight that our method can find ground state folds up to 60 beads, showing competitive results compared to other machine learning approaches in the literature.

\section{Methods}
\subsection{The HP Model}
 A fully folded protein chain can be conveniently represented on the 2D Cartesian plane. Let $\Gamma = (\gamma_0, \dots, \gamma_N) \in \{0,1\}^{N+1}$ represent an HP chain or sequence having $N+1$ beads. Here, $0$ and $1$ denote the `H' and `P' beads respectively. For a complete fold of $\Gamma$, let the Cartesian coordinates of each bead in $\Gamma$ be $(x_0,y_0), \dots, (x_N,y_N)$ respectively where $\forall i; x_i,y_i \in \mathbb{Z}$. A fold requires every pair of consecutive beads to be a unit distance away from each other, either on the $x$-axis or on the $y$-axis, but not both. In other words, the following condition
\begin{equation}
    \forall i; |x_i - x_{i+1}| + |y_i - y_{i+1}| = 1
\end{equation}
is enforced. A hard constraint on this problem is that a fold must be a self-avoiding walk (SAW), meaning no two beads can have overlapping coordinates. Thus, the constraint
\begin{equation}\label{eq:SAW}
    \forall i \neq j;  (x_i,y_i) \neq (x_j,y_j) 
\end{equation}
must also hold true for a valid fold. To map a protein sequence $\Gamma$ to some (valid or invalid) fold, we define a sequence of moves or actions to be the solution $\bm{d} \in \{0,1,2,3\}^N$. A move $d_i \in \bm{d}$ dictates the coordinates of bead $\gamma_{i}$ in a fold given the coordinates of the previous bead $\gamma_{i-1}$ which are $(x_{i-1},y_{i-1})$. Specifically, for $1 \leq i \leq N$, the next bead position is given as follows:
\begin{equation*}
    (x_i,y_i) = 
    \begin{cases}
        (x_{i-1}-1,y_{i-1}), & d_i = 0 \\
        (x_{i-1}+1,y_{i-1}), & d_i = 1 \\
        (x_{i-1},y_{i-1}+1), & d_i = 2 \\
        (x_{i-1},y_{i-1}-1), & d_i = 3. \\
    \end{cases}
\end{equation*}

The initial position $(x_0,y_0)$ can be set to any reference coordinates, such as $(0,0)$, which we use in this work. Semantically, our chosen convention is such that moves $0,1,2,$ and $3$ corresponds to placing the current bead $\gamma_i$ to the \textit{`left of', `right of', `above',} and \textit{`below'} the previous bead $\gamma_{i-1}$ respectively on the Cartesian plane.

\begin{figure}
    \centering
    \includegraphics[width=0.6\linewidth]{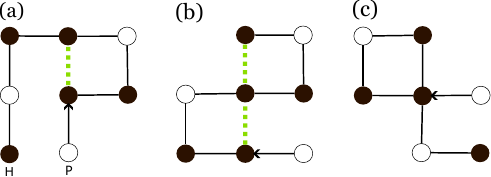}
    \caption{For the protein sequence `PHHPHHPH', encoded as $\Gamma = (1,0,0,1,0,0,1,0)$, we may have three folds as shown above. The arrow indicates the start of the fold from the first bead and the dotted green line shows the `H-H' pairs that contribute to energy. \textbf{(a)} $\bm{d}=(2,1,2,0,0,3,3)$ and $E(\bm{d})=-1$. \textbf{(b)} $\bm{d}=(0,0,2,1,1,2,0)$ and $E(\bm{d})=-2$. \textbf{(c)} $\bm{d}=(0,0,2,1,3,3,1)$ and since $\bm{d}$ breaks the self-avoiding walk constraint by overlapping the second and sixth beads, $E(\bm{d})=0$.}
    \label{fig:folds}
\end{figure}

The energy of a fold $\bm{d}$ given by $E(\bm{d})$ is defined as the negative of the number of neighboring or adjacent `H-H' pairs in the folding space that are not consecutive in the protein sequence itself, as illustrated in Fig.~\ref{fig:folds}. We can denote this number as $N_{\text{HH}}$. To formulate $E(\bm{d})$ mathematically, we first define the function $M: \mathbb{Z} \times \mathbb{Z} \rightarrow \{0,1\}$ as
\begin{equation}
        M(x,y) =
    \begin{cases}
        1, \text{   if $(x,y)$ is occupied by an $H$ bead} \\
        0, \text{   otherwise}. 
    \end{cases}
\end{equation}
Using the function $M$, we define the energy of a \textit{valid} protein sequence fold $\boldsymbol{d}$ as
\begin{equation}
    E(\boldsymbol{d}) \equiv -N_{\text{HH}}
    = \frac{1}{2} \sum_{i=0}^{N} \Bigl( (1 - \gamma_{i})( A_i -\widehat{M}_i) \Bigr),
    \label{eq:energy}
\end{equation}
where
\begin{equation*}
    \widehat{M}_i = M(x_{i-1},y_{i}) + M(x_{i+1},y_{i})
    + M(x_{i},y_{i-1}) + M(x_{i},y_{i+1}) 
\end{equation*}
and 
\begin{equation*}
    A_i = (1-\gamma_{i-1}) + (1-\gamma_{i+1}).
\end{equation*}
Here, the energy function checks all four neighboring coordinates of $\gamma_i$ in the fold. An energy contribution of $-1$ is added in proportion to the number of neighboring `H' beads through the term proportional to $\widehat{M_i}$. However, for all beads, --- except for the first and the last --- two of the four neighboring beads $\gamma_{i-1}$ and $\gamma_{i+1}$ cannot contribute to the energy as they are consecutive to $\gamma_i$ in the protein sequence. Therefore, we subtract these contributions by adding the term proportional to $A_i$. For the boundary cases of the first and the last bead, we use $\gamma_{-1} = \gamma_{N+1} = 1$. Note that the double counting of `H-H' adjacent pairs is taken into account by dividing by a factor of 2. Lastly, if a fold $\boldsymbol{d}$ is \textit{invalid}, its energy is set to $0$ by default.

\begin{figure}[h]
    \centering\includegraphics[width=\linewidth]{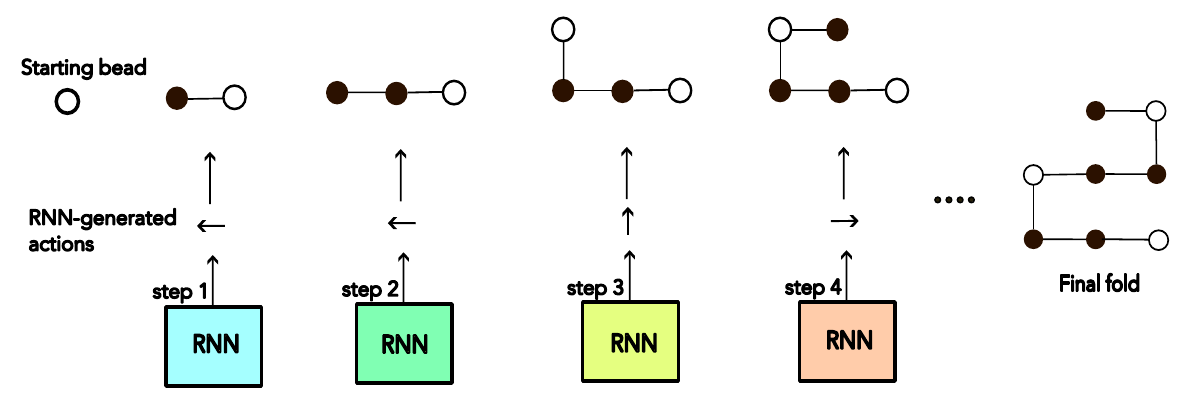}
    \caption{The sequence `PHHPHHPH' is being folded one RNN-generated move or action $d$ at a time, for a total of $N=7$ steps. We see how each action positions the next bead relative to the previous bead in the partial fold. Note the different colors on the RNN blocks to denote position-wise dedicated parameters instead of the same set of shared parameters for all RNN blocks as specified in Sec.~\ref{sec:RNN}.}
    \label{fig:folding_steps}
\end{figure}

\subsection{Variational Annealing}

Given the NP-completeness of the folding process~\cite{fraenkel_complexity_1993,berger_protein_1998}, it is natural to treat the problem of finding the optimal solution as a combinatorial problem. This setting motivates our variational learning approach which consists of sampling solutions from a distribution $P_{\boldsymbol{\theta}}$ characterized by a probabilistic model with parameters $\boldsymbol{\theta}$. In particular, we want $P_{\boldsymbol{\theta}}$ to approximate the Boltzmann distribution at a given temperature $T$ \cite{wu2019solving}. To train the model parameters $\boldsymbol{\theta}$, we use the variational free energy~\cite{wu2019solving, VNA2021}
\begin{equation}
    F_{\boldsymbol{\theta}}(T) = \langle E \rangle - T S_{\boldsymbol{\theta}},
\end{equation}
where $\langle . \rangle$ denotes the expectation of a random variable over the RNN probability distributions $P_{\boldsymbol{\theta}}$, and $S_{\boldsymbol{\theta}} = \langle - \log(P_{\boldsymbol{\theta}}) \rangle =  \sum_{\bm{d}} P_{\boldsymbol{\theta}}(\bm{d}) [- \log(P_{\boldsymbol{\theta}}(\bm{d}))]$ is the Shannon entropy. Here, $F_{\boldsymbol{\theta}}(T)$ computes the free energy over the entire state space $\boldsymbol{d} \in \{0,1,2,3\}^N$ which is intractable to compute exactly. To go around this challenge, we estimate $F_{\boldsymbol{\theta}}(T)$ by drawing $M$ independent samples $\{\boldsymbol{d^{(i)}}\}_{i=1}^{M}$ from the RNN distribution $P_{\boldsymbol{\theta}}$ and we compute an estimate of the free energy as follows:
\begin{equation}\label{eq:FreeEnergyGeneral}
F_{\boldsymbol{\theta}}(T) \approx \frac{1}{M} \sum_{i=1}^{M} \Bigl( E(\boldsymbol{d}^{(i)}) + T\log(P_{\boldsymbol{\theta}}(\boldsymbol{d}^{(i)})) \Bigr).
\end{equation}
Lastly, we note that by virtue of autoregressive sampling, the probability of sampling a fold $\boldsymbol{d} \sim  P_{\boldsymbol{\theta}}$ is given by the probability chain rule
\begin{equation}\label{eq:chain_rule}
    P_{\boldsymbol{\theta}}(\boldsymbol{d}) = \prod_{i=1}^{N} P_{\boldsymbol{\theta}}(d_i|d_1, \dots, d_{i-1}),
\end{equation}
where $P_{\boldsymbol{\theta}}(\boldsymbol{d})$ is the joint probability obtained by the product of all the conditional probabilities. Note that $P(d_i | d_{j<i})$ is the conditional probability of sampling the $i^{\text{th}}$ move $d_i$ given the realizations of all previous displacements $\{d_j\}_{j=1}^{i-1}$.

In Eq.~\eqref{eq:FreeEnergyGeneral}, the term $T \log(P_{\boldsymbol{\theta}} (\boldsymbol{d}^{(i)}))$ can be seen as an entropy regularization term weighted by temperature $T$. We use this entropy regularization to mitigate the effects of local minima in the optimization landscape~\cite{VNA2021} and also to avoid mode collapse~\cite{wu2019solving}. $T$ is annealed or cooled from a starting temperature $T_0$ to a final temperature $0$ with the possibility of varying curvatures in its descent depending on the annealing schedule – ranging from a steady, linear decay to a faster, nonlinear decay that follows the curvature of the inverse function for example. Selecting an annealing schedule is a design choice that dictates how fast $T$ decays during the different stages of the annealing process. With these schedules, entropy regularization can be seen as a mechanism that encourages exploration in the folds landscape at high temperatures before exploitation by targeting the low energy folds near zero temperature.

\subsection{Probabilistic Model}
\label{sec:RNN}
To model the probability distribution from which folds are sampled from $P_{\boldsymbol{\theta}}(\boldsymbol{d})$, we use a Dilated RNN architecture~\cite{chang2017dilated}. The motivation behind using an RNN architecture is to enable autoregressive sampling, which is a form of perfect sampling that mitigates the challenges of Markov Chain sampling schemes of other neural network architectures~\cite{Hinton2012,du2020implicitgenerationgeneralizationenergybased}. Furthermore, unlike the vanilla RNN model, Dilated RNNs have long recurrent skip connections that allow for the direct propagation of hidden state information from earlier inputs $\bm{x}_i$ to be utilized in later stages of the folding process. This property is particularly useful in the context of folding as we may want to put an `H' bead $\gamma_i$ adjacent to an `H' bead $\gamma_j$ where $i-j$ is large. These long-term dependencies benefit from the introduced dilated recurrent connections~\cite{chang2017dilated}. Fig~\ref{fig:folding_steps} illustrates the step-by-step procedure for generating a complete fold from the starting bead by the Dilated RNN.

\begin{figure}[h]
    \centering\includegraphics[width=0.5\linewidth]{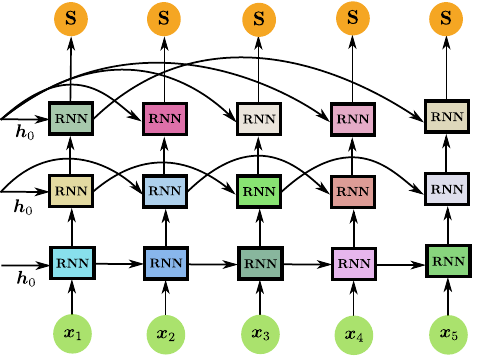}
    \caption{An illustration of a Dilated RNN architecture with $\lceil \log_2(N) \rceil$ layers, where $N$ represents the system size. The architecture incorporates longer recurrent connections to address long-range interactions in the HP lattice protein folding model. The use of distinct colors indicates the absence of weight sharing across different RNN units in the layers. $\bm{h}_0$ is an initial hidden state initialized as a zero vector and $\bm{x}_n$ are the inputs which include information about the previous move $\bm{d}_{n-1}$ and the nature of the bead $\bm{q}_n$ to be added to the chain at step $n$.}
    \label{fig:DRNN}
\end{figure}

The Dilated RNN architecture is composed of multiple layers of RNN cells stacked on top of each other as illustrated in Fig. \ref{fig:DRNN}. As a design choice, we use $L = \lceil \log_2(N) \rceil$ layers, and each layer has $N$ RNN cells~\cite{VNA2021, Khandoker_2023}. Each RNN cell is indexed by layer $l$ where $1 \leq l \leq L$ and column $n$ where $1 \leq n \leq N$. Additionally, we choose each of the $L \times N$ RNN cells to have its own unique set of dedicated parameters as opposed to the traditional practice of using multiple RNN cells sharing the same set of parameters. We use non-weight sharing to take account of the randomness of the chain sequences $\Gamma$ in a similar spirit to previous work~\cite{Khandoker_2023, VNA2021}. Parameter notations are as follows: an RNN cell at layer $l$ and column $n$ has the set of weight parameters $W_n^{(l)}$ and $U_n^{(l)}$, bias vector parameter $\boldsymbol{b}_n^{(l)}$, and an associated hidden state vector $\boldsymbol{h}_n^{(l)}$. The hidden state is computed as
\begin{equation}
\boldsymbol{h}_n^{(l)} = \texttt{tanh}(W_n^{(l)}\boldsymbol{h}_{\text{max}(0,n-2^{l-1})}^{(l)} + U_n^{(l)} \boldsymbol{h}_{n}^{(l-1)} + \boldsymbol{b}_n^{(l)})\label{eq:hn_dilated}.
\end{equation}
Here, $\boldsymbol{x}_{n-1}$ is the input (to the first layer of the Dilated RNN stack) that is a concatenation of the one-hot encoding of the protein bead for which we want to sample a fold for $\boldsymbol{q}_n$ and the one-hot encoding of the previously sampled output fold $\boldsymbol{d}_{n-1}$. More concretely, $\boldsymbol{q}_n$ is a one-hot encoding vector of $\{0,1\}$ where $0$ represents `H' and $1$ represents `P', and $\boldsymbol{d}_n$ is the one-hot encoding vector of integer $d_n \in \{0,1,2,3\}$. Using these two vectors, we can construct the input $\boldsymbol{x}_{n-1} = [\boldsymbol{q}_{n} \frown \boldsymbol{d}_{n-1}]$ where $\frown$ is the concatenation operation. Note that the initializations of the hidden state are defined as $\boldsymbol{h}_n^0 = \boldsymbol{x}_{n-1}$ and $\boldsymbol{h}_0^l = [\boldsymbol{q}_{0} \frown \boldsymbol{0}]$.

To get the output after the last layer of RNN cells, the $n^{\text{th}}$ hidden state of the last layer $\boldsymbol{h}_{n}^{(L)}$ is fed into the respective dense layer having weight $V_n$ and bias $\boldsymbol{c}_n$. As a result, we get the probability distribution for all the four folding directions of the $n^{\text{th}}$ displacement. This probability distribution vector $\boldsymbol{P}^u_n \in [0,1]^4$ is computed as
\begin{equation}
\boldsymbol{P}^u_n = \texttt{softmax}(V_n\boldsymbol{h}_n + \boldsymbol{c}_n)\label{eq:condprob_site}.
\end{equation}
Finally, the $n^{\text{th}}$ move is sampled from this distribution with the unmasked conditional probability
\begin{equation}
P^u(d_n | d_{i<n}) = \boldsymbol{P}^u_n \cdot \boldsymbol{d}_n,
\end{equation}
where $\cdot$ is the dot product. All $N$ conditional probabilities are computed sequentially to compute the joint probability of the list of displacements $\boldsymbol{d}$ in Eq.~\eqref{eq:chain_rule}.

Finally, we would like to highlight that the computational complexity of our model for sampling a configuration or for a forward pass is $\mathcal{O}(N \log(N))$, which makes our RNN model more scalable compared to traditional Transformers~\cite{vaswani2017attention}. However, at an increased cost of complexity, transformers may better capture long-range interactions.

\begin{figure}[h]
    \centering
    \includegraphics[width=0.7\linewidth]{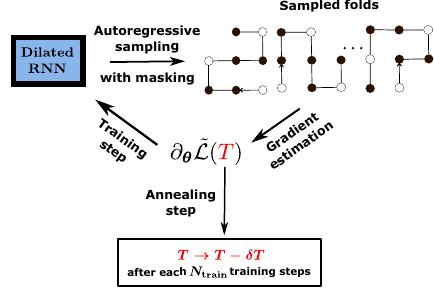}
    \caption{A diagram depicting a dilated recurrent neural network (RNN) training process applied to the HP lattice model. The RNN samples folds from its parameterized probability distribution with a mask to generate valid folds. These folds are key to estimating the gradients, which are later used to update the parameters of the Dilated RNN. Note that $T \rightarrow T-\delta T$ denotes temperature annealing i.e. decreasing temperature $T$ after every $N_{\rm train} = 5$ training steps.}
    \label{fig:workflow}
\end{figure}

\subsection{Masking and Upper Bound Optimization}\label{sec:projection}

Sampling from the valid space of folds is crucial to stabilizing training the RNN architecture and getting low-energy folds. To ensure sampling of the RNN is within the valid space, we mask the invalid moves in each RNN conditional probability $P^{u}_{\bm{\theta}}(\bm{d}_i|\bm{d}_{j<i})$ as follows:

\begin{itemize}
    \item If a direction $d_i$ is invalid, then \( \log \left( P^{u}_{\bm{\theta}}(\bm{d}_i | \bm{d}_{j<i}) \right) \) is set to $-\infty$.
    \item We renormalize the four-dimensional log conditional probability by applying the log-softmax activation and we denote it as  \( \log \left( P_{\bm{\theta}}(. | \bm{d}_{j<i}) \right) \).
\end{itemize}
Note that there are dead-end folds, where at a certain step all the local moves are invalid. In this case, the masking procedure is forced to choose a random invalid direction which results in an invalid fold. In this scenario, we discourage the RNN from generating such folds by forcing an energy penalty $E = 0$. Note that the masking step is similar in spirit to other projection schemes explored in the literature~\cite {solozabal2020constrainedcombinatorialoptimizationreinforcement, Hibat_Allah_2020}.

Although we sample a fold $\bm{d}$ from the valid fold space, we use the unmasked RNN probability $P_{\bm{\theta}}^u(\bm{d})$ for training, which amounts to training an upper bound of the free energy. This choice allows us to stabilize training and obtain lower energy folds as demonstrated in the Results section. To show the upper bound claim, let us focus on the fake loss function used to estimate the gradients of the true free energy:
\begin{equation}
   \mathcal{L}(T) = \sum_{\bm{d}} P^{\perp}_{\boldsymbol{\theta}}(\bm{d}) \log(P_{\boldsymbol{\theta}}(\bm{d})) \left( E(\boldsymbol{d}) + T \log(P^{\perp}_{\boldsymbol{\theta}}(\boldsymbol{d})) \right),
   \label{eq:fake_loss}
\end{equation}
such that $P^{\perp}_{\boldsymbol{\theta}}$ is the masked RNN probability with a stop gradient assignment $\perp$~\cite{VNA2021,Zhang_2023}. Minimizing $\mathcal{L}(T)$ corresponds to the REINFORCE method~\cite{Sutton1999} with a vanilla policy gradient rule~\cite{Shakir2020, grooten2022vanillapolicygradientoverlooked} supplemented with an entropy term~\cite{wu2019solving}. To train our Dilated RNNs, we use the following cost function:
\begin{equation}
    \tilde{\mathcal{L}}(T) = \sum_{\bm{d}} P^{\perp}_{\boldsymbol{\theta}}(\bm{d}) \log(P^{u }_{\boldsymbol{\theta}}(\bm{d})) \left( E(\boldsymbol{d}) + T \log(P^{u\perp}_{\boldsymbol{\theta}}(\boldsymbol{d})) \right)
\label{eq:fake_loss_upper_bound}
\end{equation}
where $P^{u}_{\boldsymbol{\theta}}$ is the unmasked RNN probability. The following inequality $$\mathcal{L}(T) \leq \tilde{\mathcal{L}}(T)$$follows from the observation $$P_{\boldsymbol{\theta}}(\bm{d}) \geq P^{u }_{\boldsymbol{\theta}}(\bm{d})$$
for all possible valid folds $\bm{d}$, which implies that:
\begin{align*}
    E(\boldsymbol{d}) \log P_{\boldsymbol{\theta}}(\bm{d})  &\leq E(\boldsymbol{d}) \log P^{u}_{\boldsymbol{\theta}}(\bm{d})  \\
    \log^2 P_{\boldsymbol{\theta}}(\bm{d}) &\leq \log^2 P^{u }_{\boldsymbol{\theta}}(\bm{d}).
\end{align*}
The first inequality follows from the fact that $E(\bm{d})\leq 0$ for all possible folds $\bm{d} \in \{0,1,2,3\}^N$. Training using the upper bound loss function $\tilde{\mathcal{L}}(T)$ follows a similar spirit to training variational autoencoders~\cite{kingma2022autoencodingvariationalbayes} and diffusion models~\cite{sohldickstein2015deepunsupervisedlearningusing, ho2020denoisingdiffusionprobabilisticmodels} with upper bound loss functions. The gradient of the free energy upper bound is given as:
\begin{equation*}
    \partial_{\boldsymbol{\theta}}  \tilde{\mathcal{L}}(T) =
        \sum_{\bm{d}} P^{\perp}_{\boldsymbol{\theta}}(\bm{d}) \partial_{\boldsymbol{\theta}}\log(P^{u }_{\boldsymbol{\theta}}(\bm{d})) \left( E(\boldsymbol{d}) + T \log(P^{u\perp}_{\boldsymbol{\theta}}(\boldsymbol{d})) \right),
\end{equation*}
which can be estimated by sampling $M$  folds autoregressively from the RNN as follows:
\begin{equation}\label{eq:LossFinal}
    \partial_{\boldsymbol{\theta}} \tilde{\mathcal{L}}(T) \approx 
        \frac{1}{M} \sum_{\bm{d} \sim P_{\boldsymbol{\theta}}} \left( \partial_{\boldsymbol{\theta}} \log P^{u}_{\boldsymbol{\theta}}(\bm{d}) \right) \left( \overline{E(\boldsymbol{d})} + T \overline{\log(P^{u}_{\boldsymbol{\theta}}(\boldsymbol{d}))} \right).
\end{equation}
Note that we used the notation $\overline{O(\boldsymbol{d})} \equiv O(\boldsymbol{d}) - \langle O \rangle$, where subtracting the average of the energies and log probabilities was shown to reduce the variance in the gradients as a control variate method~\cite{Hibat_Allah_2020, Shakir2020}.

\section{Experiments \& Results}

An overview of the training scheme of our RNN is presented in Algorithm~\ref{alg:algo} and also illustrated in Fig.~\ref{fig:workflow}. In our method, we use a linear schedule for the annealing process which is characterized by Line 3 in the pseudocode. Further hyperparameter details are provided in Appendix~\ref{app:hyperparameters}, Tab.~\ref{tab:results_appendix}.
\begin{algorithm}[th]
    \caption{Variational Annealing Training}
    \label{alg:algo}
    \begin{algorithmic}[1]
        \State \textbf{Required:} $\Gamma, T_0, N_\text{anneal}, N_\text{train}, M$ 
        \For{$N_\text{step} \in \{0, \dots, N_\text{anneal}\}$}
            \State $T \gets T_0(1 - N_\text{step} / N_\text{anneal})$
            \For{$i \in \{1, \dots, N_\text{train}\}$}
                \State Sample $M$ independent folds of protein $\Gamma$ from the RNN.
                \State Compute the gradients $\partial_{\boldsymbol{\theta}} \tilde{\mathcal{L}}(T)$ (Eq. \ref{eq:LossFinal}).
                \State Update the RNN parameters using the Adam optimizer~\cite{KingBa15}.
            \EndFor
        \EndFor
    \end{algorithmic}
\end{algorithm}

\begin{figure}
    \centering
    \includegraphics[width=\linewidth]{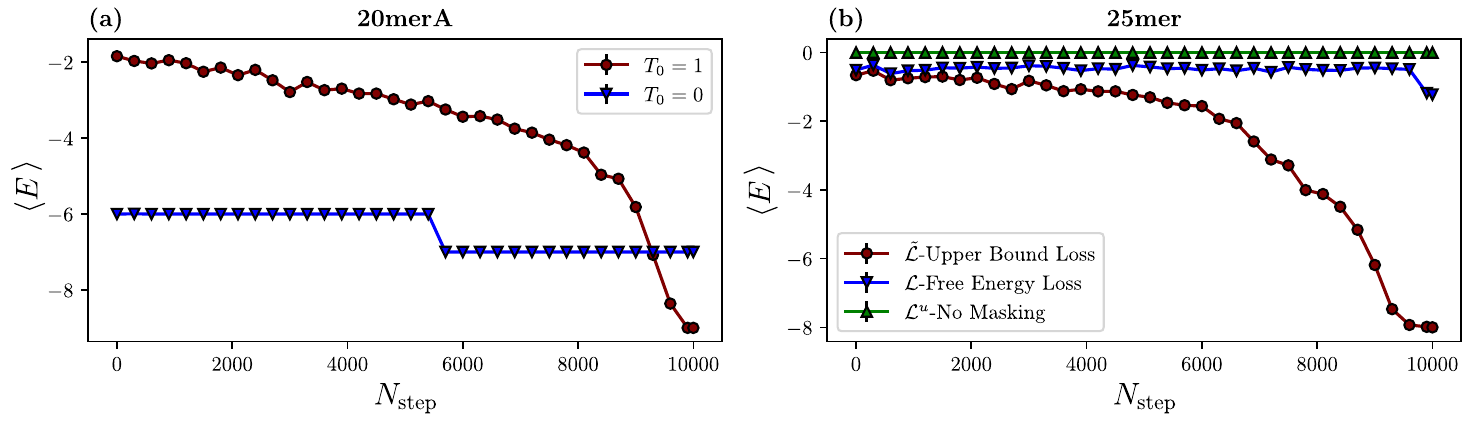}
    \caption{Figures of the training process of the variational annealing approach with a total number of annealing steps $N_{\text{anneal}}=10000$. \textbf{(a)} Demonstrates the effect of annealing in training for the sequence 20merA of the expectation value of the energy $\langle E \rangle$ as a function of the number of temperature annealing steps $N_{\text{step}}$ where each step corresponds to $N_\text{train} = 5$ training steps at the same temperature. Note that the user-defined value $N_{\text{anneal}}$ is the maximum value that $N_{\text{step}}$ can reach. \textbf{(b)} Demonstrates the training process using masking with the upper bound loss $\mathcal{\tilde{L}}(T)$, masking with the fake free energy loss $\mathcal{L}(T)$, and no masking with the unmasked loss $\mathcal{L}^u(T)$ for the sequence 25mer.}
    \label{fig:anneal}
\end{figure}

\begin{table*}[h]
   \centering
   \begin{tabularx}{\textwidth}{@{}cccc@{}}
   \toprule
   HP Sequence &  & Length & $E(\boldsymbol{d^*})$ \\
   \midrule
   20merA  & HPHPPHHPHPPHPHHPPHPH & 20 & $-9$ \\
   20merB  & HHHPPHPHPHPPHPHPHPPH & 20 & $-10$ \\
   24mer   & HHPPHPPHPPHPPHPPHPPHPPHH & 24 & $-9$ \\
   25mer   & PPHPPHHPPPPHHPPPPHHPPPPHH & 25 & $-8$ \\
   36mer   & PPPHHPPHHPPPPPHHHHHHHPPHHPPPPHHPPHPP & 36 & $-14$ \\
   48mer   & PPHPPHHPPHHPPPPPHHHHHHHHHHPPPPPPHHPP- & 48 & $-23$ \\
           & HHPPHPPHHHHH \\
   50mer   & HHPHPHPHPHHHHPHPPPHPPPHPPPPHPPPHPPPH- & 50 & $-21$ \\
           & PHHHHPHPHPHPHH \\
   60mer   & PPHHHPHHHHHHHHPPPHHHHHHHHHHPHPPPHHHH- & 60 & $-36$ \\
           & HHHHHHHHPPPPHHHHHHPHHPHP \\
   \bottomrule
   \end{tabularx}
   \caption{HP sequences from the Istrail Benchmark with their best known energies $E(\boldsymbol{d^*})$.}
\label{tab:benchmark}
\end{table*}

\begin{figure}
    \centering
    \includegraphics[width=12cm, height=4.5cm]{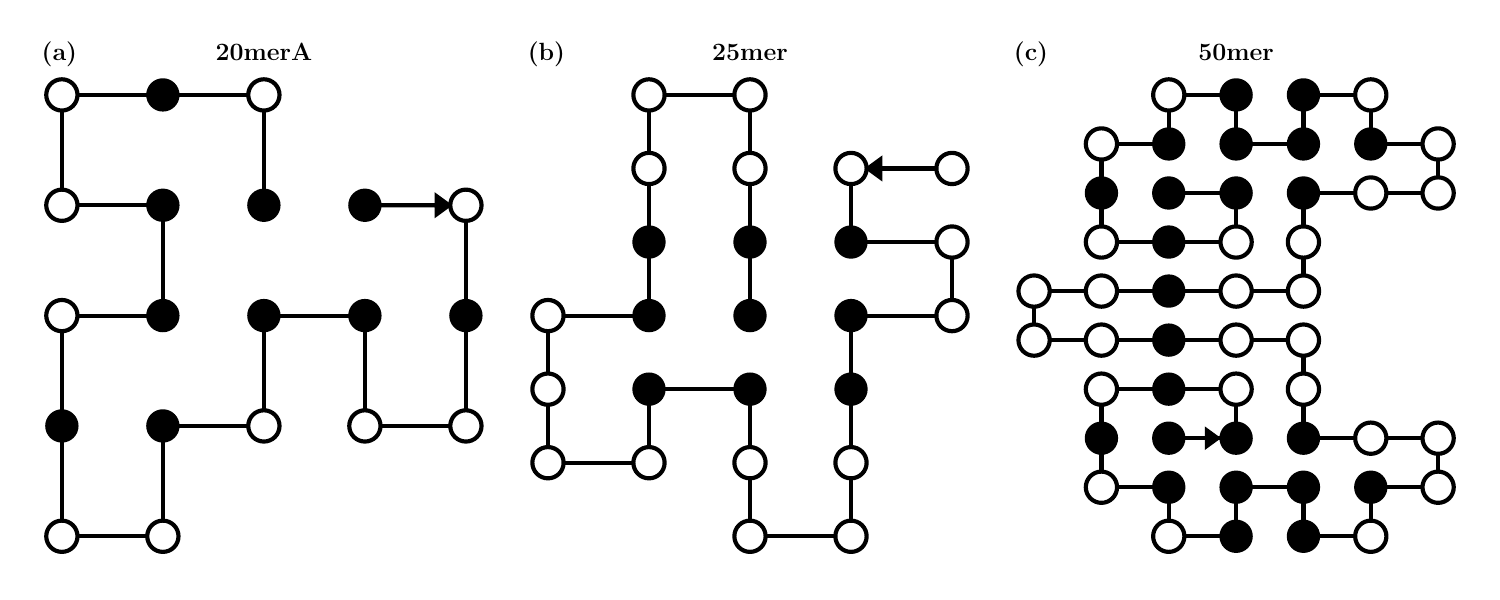}
    \caption{Some representative, optimal folds found by the VNA method for HP sequence \textbf{(a)} 20merA, \textbf{(b)} 25mer, and \textbf{(c)} 50mer. The solid and hollow circles represent `H' and `P' beads respectively. The arrow indicates the start of the fold.}
    \label{fig:optimal_folds}
\end{figure}

\begin{table*}[h]
\centering
    \begin{tabularx}{\textwidth}{@{\space}>{\centering\arraybackslash}X>{\centering\arraybackslash}X>{\centering\arraybackslash}X>{\centering\arraybackslash}X>{\centering\arraybackslash}X>{\centering\arraybackslash}X>{\centering\arraybackslash}X@{\space}}
    \toprule
    HP Sequence & $E(\boldsymbol{d^*})$ & Folding-Zero & AlphaGo-Zero & DRL & DQN-LSTM & Variational Annealing (\textbf{ours}) \\
    \midrule
    20merA & \textbf{-9} & \textbf{-9} & -8 & -6 & \textbf{-9} & \textbf{-9} \\
    20merB & \textbf{-10} & - & -9 & -8 & \textbf{-10} & \textbf{-10} \\
    24mer & \textbf{-9} & -8 & -8 & -6 & \textbf{-9} & \textbf{-9} \\
    25mer & \textbf{-8}  & -7 & -7 & - & \textbf{-8} & \textbf{-8} \\
    36mer & \textbf{-14} & -13 & -13 & -& \textbf{-14} & \textbf{-14} \\
    48mer & \textbf{-23} & -18 & - & - & \textbf{-23} & \textbf{-23}\\
    50mer & \textbf{-21} & -18 & - & - & \textbf{-21} & \textbf{-21} \\
    60mer & \textbf{-36} & - & - & - & - & \textbf{-36} \\
    \bottomrule
    \end{tabularx}
    \caption{Lowest energy found by the following methods in order: Folding-Zero \cite{li2018foldingzeroproteinfoldingscratch}, AlphaGo Zero with pretraining \cite{Yu2020DeepRL}, (best results among various methods of) Deep Reinforcement Learning \cite{Yu2020DeepRL}, Deep Q-Network using Long Short Term Memory (DQN-LSTM) \cite{Yang_2023}, and our Variational Annealing method. Bold font represents the lowest energy of the corresponding sequence. `-' means energy was not reported by the respective authors.}
\label{tab:results}
\end{table*}

\subsection{Annealing Ablation}
To demonstrate the empirical value of annealing, we show in Fig.~\ref{fig:anneal}(a) a comparison between the expectation values of the energy $\langle E \rangle = \sum_{\bm{d}} P_{\boldsymbol{\theta}}(\bm{d}) E(\bm{d})$ between variational learning with annealing (by setting $T_0=1$) and without annealing (by setting $T_0=0$). Recall that this expectation is estimated by taking the average of $M$ independent samples to get $\langle E \rangle \approx (1/M) \sum_{i=1}^{M}E(\bm{d}_i)$. Our findings show that starting at a non-zero temperature during annealing allows our dilated RNN to find lower energy folds compared to a plain optimization scheme that does not include entropy regularization.

\subsection{Masking Ablation}
Additionally, we empirically support the use of masking in tandem with our derived upper bound loss based on the ablation study in the training process as shown in Fig.~\ref{fig:anneal}(b). We compare the results of masking with the upper bound loss $\mathcal{\tilde{L}}(T)$, masking with the fake free energy loss $\mathcal{L}(T)$, and no masking with the unmasked loss
\begin{equation*}
        \mathcal{L}^u(T) =
        \sum_{\bm{d}} P^{u\perp}_{\boldsymbol{\theta}}(\bm{d}) \log(P^{u }_{\boldsymbol{\theta}}(\bm{d})) \left( E(\boldsymbol{d}) + T \log(P^{u\perp}_{\boldsymbol{\theta}}(\boldsymbol{d})) \right).
\end{equation*}
From the results of Fig.~\ref{fig:anneal}(b), it is clear that enforcing the self-avoiding walk constraint Eq.~\eqref{eq:SAW} by masking is a crucial step for enhancing the training of our RNN model. More specifically, both $\mathcal{\tilde{L}}(T)$ and $\mathcal{L}(T)$ have lower values compared to $\mathcal{L}^u(T)$ throughout annealing. Furthermore, we also observe that optimizing the free energy upper bound loss $\mathcal{\tilde{L}}(T)$ can result in significantly lower energies compared to the free energy fake loss function $\mathcal{L}(T)$ as illustrated in Fig.~\ref{fig:anneal}(b). We believe that masking the RNN probability $P^u_{\boldsymbol{\theta}}$ to obtain $P_{\boldsymbol{\theta}}$ is constraining the optimization of $\mathcal{L}(T)$ when we apply the gradient on the $\log P_{\boldsymbol{\theta}}$ terms in Eq.~\eqref{eq:fake_loss}. In contrast, the loss upper bound $\mathcal{\tilde{L}}(T)$ in Eq.~\eqref{eq:fake_loss_upper_bound} circumvents this limitation where the gradients are applied on $\log P^u_{\boldsymbol{\theta}}$ terms (Eq.~\eqref{eq:fake_loss_upper_bound}) that are not constrained by projection steps.

\begin{figure}[h]
    \centering
    \includegraphics[width=\linewidth]{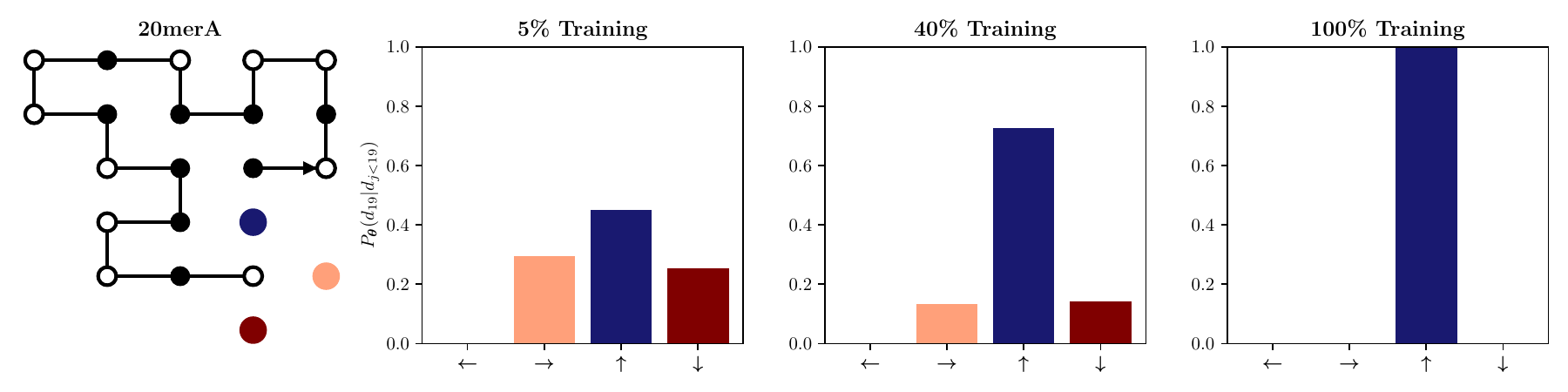}
    \caption{Conditional probability distribution over the four moves for the last step $d_{19}$ of the 20merA sequence at different stages of the training process. Notice that the conditional probability of the invalid \textit{`left'} move is always $0$ and by the end of training, all the probability mass is on the optimal \textit{`up'} move.}
    \label{fig:actions}
\end{figure}

\subsection{Benchmarks}
To compare our results with the machine learning literature, we present in Tab.~\ref{tab:benchmark} a list of HP sequences with lengths from $20$ to $60$ and their optimal energies that are often used to benchmark HP folding algorithms~\cite{Yang_2023}. We use these HP sequences to benchmark our algorithm and compare with recent machine learning methods on the criteria of the lowest energy fold found. The results are reported in Tab.~\ref{tab:results} and some optimal folds that our method finds are shown in Fig.~\ref{fig:optimal_folds}. Our findings demonstrate that our variational annealing method with dilated RNNs can find the optimal fold for all the sequences up to $60$ beads. We also highlight that our method performs better than previous studies that take inspiration from the AlphaGo Zero algorithm \cite{li2018foldingzeroproteinfoldingscratch,Yu2020DeepRL}. Additionally, we note that we only use $d_h = 50$ as the size of the hidden state in our dilated RNN compared to the LSTM used in Ref.~\cite{Yang_2023} with $d_h \in \{256, 512\}$. This observation highlights the computational efficiency of our method. Most importantly, we identify the optimal fold for the 60-bead configurations, representing a notable improvement over the machine learning optimization techniques reported in this paper (see Tab.~\ref{tab:results}). We performed thorough experiments on chains up to a reasonable size of 60 residues, but note that experiments on larger chains can also be carried out, albeit at higher computational costs. Furthermore, we report an example of the evolution of the conditional probability of the actions made by the RNN in Fig.~\ref{fig:actions} for the 20merA sequence. Here the confidence in making an action that increases the number of H-H interactions is increasing with more training steps, indicating the RNN is indeed learning to maximize the number of H-H connections. The low-energy states of the HP model which give rise to the structures of the folded proteins (see Fig.~\ref{fig:optimal_folds}) consist of a core of H beads that exhibit high compactness. This confirms known results regarding the folded states of HP sequences~\cite{lau1989lattice}. Consequently, this suggests that the variational annealing method could help describe the principles underlying lattice protein folding.

\begin{figure}[h]
    \centering\includegraphics[width=\linewidth]{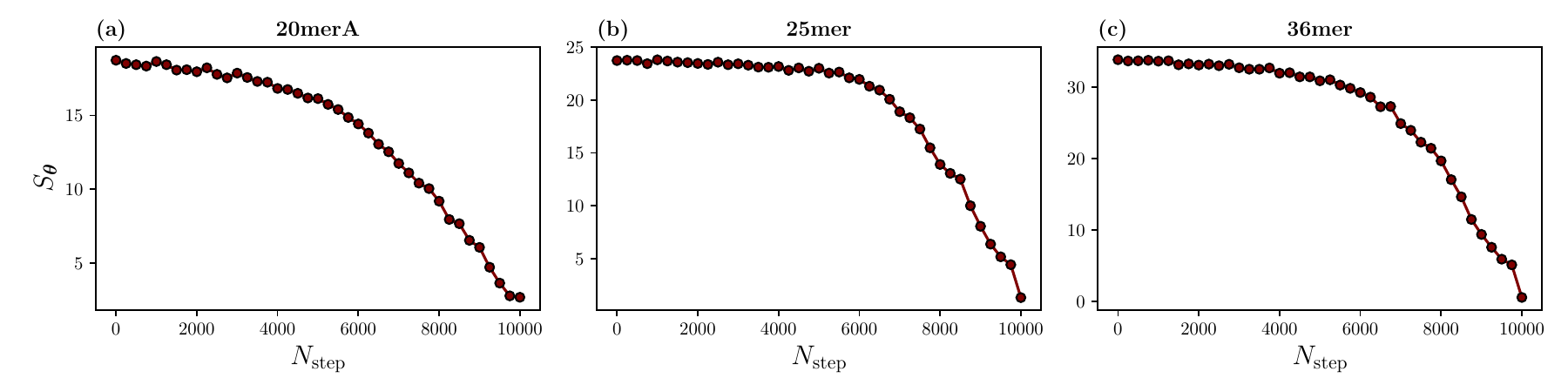}
    \caption{Figures of the Shannon entropies $S_{\boldsymbol{\theta}}$ during training at every annealing stage spanning $N_{\text{annealing}}=10000$ steps on the sequences \textbf{(a}) 20merA, \textbf{(b)} 25mer, and \textbf{(c)} 36mer.}
    \label{fig:entropyasd}
\end{figure}

Finally, we highlight that our RNN model finds diverse solutions at the end of annealing. In particular, Fig.~\ref{fig:entropyasd} shows the evolution of the Shannon entropy of the RNN as a function of the number of annealing steps $N_{\rm step}$. By the end of annealing, we find non-zero final entropy values of $2.69,1.33, 0.56$ for the 20merA, 25mer, and 36mer protein chains respectively. This observation highlights the ability of our RNN model to find a diverse set of solutions. 

\section{Conclusion}
In this paper, we introduce a novel upper-bound training scheme with masking to find the lowest energy fold of the 2D HP lattice protein folding. Using this scheme and by supplementing Dilated RNNs with annealing through temperature-like fluctuations, we find that our method can find the optimal folds of prototypical lattice folding benchmarks with system sizes up to 60 beads. We demonstrate that it is possible to mask moves that lead to invalid folds without compromising the autoregressive sampling feature of RNNs. We also devise a free energy upper bound loss function that enhances the trainability of RNNs that are prone to get compromised by masking the RNN probabilities. We note that there is still room for exploring variational annealing with other neural network architectures, such as Transformers~\cite{vaswani2017attention} on the lattice folding problem and beyond.

To motivate future lines of inquiry, we want to highlight that our scheme is generalizable to folding HP sequences in three spatial dimensions with no significant increase in computational complexity in the forward pass with respect to input size $N$. The 3D model introduces two additional moves to the base set of four moves, which are the \textit{`forward'} and \textit{`backward'} moves on the $z$-axis. This setting can be easily implemented by increasing the size of RNN output vector $\boldsymbol{d}_i$ to six and the input vector $\bm{x}_i$ to eight. The energy function will then take into account six neighbors instead of four.
Our method can also generalize to other lattice protein folding models with more than two alphabets such as the 20-letter Miyazawa-Jernigan model~\cite{miyazawa_residue_1996} by enlarging the sizes of the one-hot inputs $\bm{x}_i$ without compromising inference speed. Although our scheme extends implementation to 3D and larger-alphabet models, varying model dynamics may pose unique learning challenges. Investigating our method's interplay with a family of close-knit folding problems will help distinguish generalizable principles from those needing tailored adjustments.

We also expect inference time to be reduced by introducing an encoder network to enable just-in-time inference of low-energy folds~\cite{sanokowski2023variationalannealinggraphscombinatorial}.
Lastly, for a broader scope, we also believe that our scheme could lead to a promising machine learning-based solution to a wide class of constrained combinatorial optimization problems such as graph coloring and the traveling salesman problem.

\section*{Acknowledgments}
EMI and MH acknowledge support from the Natural Sciences and Engineering Research Council of Canada (NSERC) and the Perimeter Institute for Theoretical Physics. Research at Perimeter Institute is supported in part by the Government of Canada through the Department of Innovation, Science and Economic Development Canada and by the Province of Ontario through the Ministry of Economic Development, Job Creation and Trade. SAK also wants to thank Caleb Schultz Kisby for holding valuable discussions about the project. Computer simulations were made possible thanks to the Digital Research Alliance of Canada cluster.

\section*{Code availability}
Our code is publicly available on GitHub at \url{https://github.com/shoummoak/VA-HP-Lattice}.

\bibliographystyle{unsrt}
\bibliography{Refs.bib}
\newpage

\appendix
\section{Hyperparameters}
\label{app:hyperparameters}
\begin{table*}[h]
\centering
    \begin{subtable}{\textwidth}
    \centering
    \begin{tabularx}{\textwidth}{@{\space}>{\centering\arraybackslash}X>{\centering\arraybackslash}X>{\centering\arraybackslash}X>{\centering\arraybackslash}X>{\centering\arraybackslash}X>{\centering\arraybackslash}X@{\space}}
    \toprule
    HP Sequence & $T_0$ & $N_\text{warmup}$ & $N_\text{annealing}$ & $N_\text{train}$ & $M$ \\
    \midrule
    20merA & 1 & 1000 & 10,000 & 5 & 200\\
    20merB & 1 & 1000 & 10,000 & 5 & 200\\
    24mer & 1 & 1000 & 10,000 & 5 & 200\\
    25mer & 1 & 1000 & 10,000 & 5 & 200\\
    36mer & 1 & 1000 & 10,000 & 5 & 200\\
    48mer & 5 & 1000 & 10,000 & 5 & 200\\
    50mer & 5 & 1000 & 30,000 & 5 & 200\\
    60mer & 0.5 & 1000 & 40,000 & 5 & 200\\
    \bottomrule
    \end{tabularx}
    \caption{A summary of the hyperparameters using our annealing training protocol.}
    \label{tab:subtable1}
    \end{subtable}
    
    \vspace{0.5cm} 
    
    \begin{subtable}{\textwidth}
    \centering
    \begin{tabularx}{\textwidth}{@{\space}>{\centering\arraybackslash}X>{\centering\arraybackslash}X>{\centering\arraybackslash}X>{\centering\arraybackslash}X@{\space}}
    \toprule
    RNN Layers $L$ & Hidden Unit $\boldsymbol{h}$ size & Activation Function & Learning Rate \\
    \midrule
    $\lceil \log_2(N+1) \rceil$ & 50 & tanh & $10^{-4}$ \\
    \bottomrule
    \end{tabularx}
    \caption{A summary of the hyperparameters used for training the Dilated RNN architecture for all the protein sequences in Sec.~\ref{sec:RNN}.}
    \label{tab:subtable2}
    \end{subtable}
    
    \caption{In \textbf{(a)}, we introduce an additional hyperparameter $N_\text{warmup}$, which is used to perform an optional `warm start' before the annealed training process. In this warm start, we perform exactly $N_\text{warmup}$ gradient descent steps at fixed temperature $T_0$. Additionally, note that $T_0$ is the initial temperature, $M$ is the number of samples used for each training step, $N_{\rm{train}}$ is the number of training steps during each annealing step, and $N_{\rm annealing}$ is the total number of annealing steps. In \textbf{(b)}, we highlight that we use the same hidden unit size for each RNN layer.}
    \label{tab:results_appendix}
\end{table*}

\end{document}